\documentclass[traditabstract,printer]{aa}
\usepackage{txfonts}
\usepackage{graphicx}
\usepackage{subfigure}
\usepackage{color}
\usepackage{natbib}

\begin{document}

\title{
Spectroscopic confirmation of the low-latitude object FSR~1716 as an old globular cluster} 

\author{
Andreas Koch\inst{1,2} 
  \and Andrea Kunder\inst{3,4}
  \and Jennifer Wojno\inst{4}
  }
  
\authorrunning{A. Koch et al.}
\titlerunning{Spectroscopic study of the  globular cluster FSR~1716}
\offprints{A. Koch;  \email{a.koch1@lancaster.ac.uk}}
\institute{Zentrum f\"ur Astronomie der Universit\"at Heidelberg, Astronomisches Recheninstitut, M\"onchhofstr. 12, 69120 Heidelberg, Germany
  \and Department of Physics, Lancaster University, LA1 4YB, Lancaster, UK
  \and Saint Martin's University, Old Main, 5000 Abbey Way SE, Lacey, WA 98503, USA
  \and Leibniz-Institut f\"ur Astrophysik Potsdam, An der Sternwarte 16, 14482 Potsdam, Germany   
   }
\date{Received: 14 August 2017 / Accepted: 26 August 2017}
\abstract{
Star clusters are invaluable tracers of the Galactic components and the discovery and characterization of 
low-mass stellar systems can be used to appraise their prevailing disruption mechanisms and time scales. 
However, owing to the significant foreground contamination, high extinction, and still uncharted interfaces
of the underlying Milky Way components, objects at  low Galactic latitudes are notoriously 
difficult to characterize. Here, we present the first spectroscopic campaign to identify the chemodynamical 
properties of the low-latitude star cluster   FSR 1716. While its photometric age and distance are far from settled, 
the presence of RR Lyrae variables indicates a rather old cluster variety. 
  Using medium-resolution (R$\sim$10600) 
  calcium triplet (CaT) spectroscopy obtained with the wide-field, multi-fibre AAOmega instrument, we 
identified six member candidates with a mean velocity of $-30$ km\,s$^{-1}$ and  a velocity dispersion of 2.5$\pm$0.9 km\,s$^{-1}$.
The latter value implies a dynamic mass of $\sim$1.3$\times$10$^4$ M$_{\odot}$, typical of a low-mass globular cluster. 
Combined with our derived CaT metallicity of $-1.38\pm0.20$ dex, this object is finally confirmed as an old, metal-poor  globular cluster. 
}%
\keywords{Stars: abundances --- Galaxy: abundances --- Galaxy: structure --- Galaxy: disk  --- 
Galaxy: open clusters and associations: general --- globular clusters: individual: FSR1716}
\maketitle 
%
%
%
%
\section{Introduction}
Star clusters in any of the Galactic components (halo, disks, bulge) are crucial tracers of the prevailing stellar populations
at their birthplace. Thus they play an important role for disentangling the chemodynamic 
evolution of the Milky Way. 
For a unique mapping it is therefore imperative to be able to distinguish between 
low-mass, young objects \citep[e.g.,][]{VandePutte2010} 
and old globular clusters (GCs), or to diversify objects formed in situ versus those  
accreted from larger objects such as dwarf galaxies \citep{Marin-Franch2009,Law2010}. 
This becomes progressively more difficult towards the interfaces of the components, when 
distinctions between, e.g., the inner and outer halo or the disk-bulge-halo transition 
need to be sampled \citep[e.g.,][]{RecioBlanco2014,Koch2016,Koch2017ESO,Mucciarelli2017}. 

However, such a cartography becomes challenging towards the plane of the Milky Way, since the high 
extinction and confusion with disk field stars along the line of sight render such studies a pain in the plane. 
Past estimates of the completeness of GCs in the Milky Way \citep[e.g.][]{Ivanov2005}
indicated that about ten ($\pm$3) clusters are still missing towards the inner
Milky Way, and discovery of more unknown objects still lurking in the Galactic
anticenter direction would add invaluable information 
to the overall population of Galactic star clusters. Once discovered, it needs to be carefully 
assessed which Galactic component has spawned them, but
 without the proper chemical and/or kinematic information, this endeavour remains aleatoric. 

Here, we continue our efforts to characterize hitherto uncharted star clusters in the Milky Way \citep{Koch2017ESO,Koch2017Gaia}
by performing a spectroscopic study of the low-latitude ($l$=330$\degr$, $b$=$-$1.6$\degr$) object FSR~1716.
This star cluster was discovered by \citet{Froebrich2007FSR} in the  infrared     
Two Micron All Sky Survey \citep[2MASS;][]{Cutri2003}.  
Subsequent analysis of ESO NTT infrared (JHK) data indicated an age of $>$2 Gyr and a metal-poor nature of approximately 
$-1.6$ dex \citep{Froebrich2008}. As of then, no consensus had been reached whether this 
is an open or a globular cluster.

Upon re-analysis of the 2MASS data, \citet{BonattoBica2008} confirmed an old age between 7--12 Gyr and placed
FSR~1716 within 0.8--2.3 kpc of the  Solar { radius}\footnote{Note that this estimate used a value for the Solar radius of 7.2 kpc, whereas 
new measurements place this at larger distances of $\sim$8.3 kpc \citep[e.g.,][]{Reid2009,Brunthaler2011}}. 
However, in order to remain stable at that old age, this system needed to be 
an ``extreme dynamic survivor'', since the severe disruption mechanisms in the Galactic disk
would lead to strong mass loss. 
Accordingly, open clusters that old are mainly found at larger Galactocentric distances due to the decreased relentlessness of 
dynamical processes leading to their fast disruptions on  time scales well below 1 Gyr 
\citep{Lamers2005,BonattoBica2007,Friel2013}. 
Their derived very small radius ($r_t < 4$ pc), low stellar content (M$<$200 M$_{\odot}$), and Solar metallicity prompted 
\citet{BonattoBica2008} to prefer a characterization as a rare, old open cluster over a low-mass, old globular cluster. 
{ In contrast, \citet{Buckner2016} assigned this object an old age (10--12 Gyr), but at a larger distance to the Sun of 
7.3 kpc.}

Recently, FSR~1716 was identified via its RR Lyrae population in the deep, multi-epoch, near-infrared ``VISTA Variables in the V\'{\i}a L\'actea'' (VVV) survey \citep{Minniti2017}, which 
ascertained an old age of $>$10 Gyr and a low metallicity ($-1.5$ dex) from the RR-Lyrae's 
period-metallicity relation.  Accordingly, this analysis placed FSR~1716 at a distance well within the Solar circle, at 
$d_{\odot}$=7.5 kpc and  $R_{\rm GC}$=4.3 kpc.  Taken at face value, these values would argue in favour of this object 
indeed being of the globular variety. Therefore, only the addition of chemical abundance and detailed kinematic information 
will be able to { settle} the true nature of FSR~1716 and its proper place in the Milky Way galaxy. 

This paper is organized as follows: In Sect.~2 we describe our spectroscopic observations, from which we measure the stellar kinematics and assess membership with the star cluster in Sect.~3. 
Sect.~4 contains our derivation of metallicities, and finally, in Sect.~5, we summarize our findings. 
\section{Observations}
Targets for this program were selected from the 2MASS catalog, applying generous selection boxes that would, in principle, 
encompass the cluster's red giant branch (RGB) from its tip (K$_S$$\sim$9 mag) to its red clump level (at K$_S$$\sim$13.5 mag). 
Owing to its very low latitude, the line of sight towards FSR~1716 is heavily contaminated with Galactic disk stars, which 
renders an efficient target selection difficult (Fig.~1; see also Fig.~4 in \citealt{Minniti2017}). 
Similarly, the success rate of our procedure is limited by the small extent of the cluster: at the distance to FSR~1716 of $\sim$7.5 kpc, its 
core radius subtends a mere 26.5$\arcsec$ \citep{Froebrich2008}, while \citet{Minniti2017} estimate a full extent of the cluster of $\sim$3$\arcmin$. 
This compares to a field of view of the AAOmega instrument of 2$\degr$ and a minimum fibre separation of $\sim$30$\arcsec$ \citep{Sharp2006}.
\begin{figure}[b]
\centering
\includegraphics[width=1\hsize]{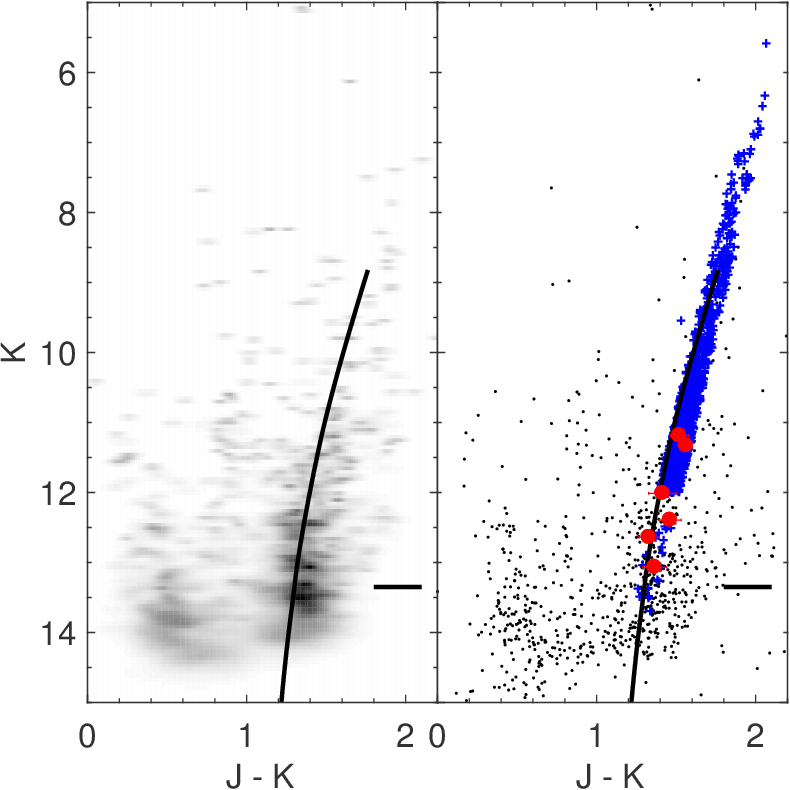}
\caption{Left panel: Hess diagram of stars within 4$\arcmin$ of the nominal cluster center using 2MASS photometry. The right panel shows the  member candidates as red points, whereas blue crosses indicate foreground stars that were used as fibre-filler targets. In either case, we highlight 
the red clump level at K$_S$=13.35 mag and a Dartmouth isochrone \citep{Dotter2008}, using the best parameters for a reddening (A$_K$=0.38 mag), distance (14.62 mag), age (10 Gyr), and 
metallicity ($-1.5$ dex) from \citet{Minniti2017}.}
\end{figure}
{ Details on the {\em bona fide} member candidates, as determined in Sect.~3, are summarized in Table~1}.
\begin{table*}[htb]
\caption{{ Photometric properties} of the FSR~1716 member candidates}              
\centering          
\begin{tabular}{cccccccc}     
\hline\hline       
  & ID & $\alpha$ & $\delta$ & $r$ & J & H & K  \\
\raisebox{1.5ex}[-1.5ex]{Star} & (2MASS) & (J2000.0) & (J2000.0) & [$\arcsec$] & [mag] &  [mag] & [mag]  \\
\hline
{ Plate~1}  \phantom{1}\#16 & J161030.35$-$534450.8 & 16:10:30.35 & $-$53:44:50.81 & \phantom{5}6.0 & 13.429 & 12.393 & 12.013  \\
{ Plate~1}  \phantom{1}\#24 & J161032.22$-$534550.3 & 16:10:32.22 & $-$53:45:50.33 & 57.8 & 12.710 & 11.547 & 11.189  \\
{ Plate~1} \#106 & J161025.78$-$534446.3 & 16:10:25.78 & $-$53:44:46.28 & 38.7 & 13.971 & 12.954 & 12.641 \\
{ Plate~2} \#266 & J161026.77$-$534407.7 & 16:10:26.77 & $-$53:44:07.73 & 56.1  & 13.858 & 12.840 & 12.395 \\
{ Plate~3}  \phantom{1}\#76 & J161029.88$-$534605.4 & 16:10:29.88 & $-$53:46:05.35 & 69.4 & 12.897 & 11.737 & 11.333 \\
{ Plate~3} \#213 & J161026.46$-$534423.2 & 16:10:26.46 & $-$53:44:23.18 & 45.4 & 14.435 & 13.426 & 13.069  \\
\hline
\hline
\end{tabular}
\end{table*}

The data presented here were taken over three nights during our six-nights' observing run in June 2017, with the AAOmega multifibre spectrograph at the 3.9 m Anglo-Australian Telescope (Siding Spring Observatory, Coonabarabran, NSW, Australia). 
We created three { plate configurations } for the Two Degree Field (2dF) fibre positioner, containing a total of 1048 science fibres and 75 sky positions. 
As before, we used a dual setup  that employed  the red 1700D grating, centered at 8600 \AA~and the blue 2500V grating, centered at 5000 \AA. This way we ensured to comprise the 
prominent calcium triplet (CaT) lines in the red. 
In the following, we will only employ the red part of the spectra { owing to the too low signal 
in our targets at the bluest wavelengths}.
The exposure times were 3$\times$1200 s for { plates} 1 and 3 and 2$\times$1500 s for { plate} 2, with overall seeing conditions of $\sim$1.1--1.4$\arcsec$.
Data reduction was carried out in a standard manner using AAO's 2dfdr pipeline { \citep{2dfdr}}, which performs quartz-flatfielding, wavelength calibration via arc-lamp exposures, sky subtraction using the dedicated sky fibres, and optimal extraction of the science spectra. 
The final wavelength range is  
8350--8800 \AA, with slight variations depending on the exact position of the spectra on the CCD. 
Our spectra reach a resolving power  
of R$\sim$10600, as measured from the width of the calibration arc lamps. 
Finally, the median signal- to-noises ratio (S/N) of our final extracted spectra is  
30 px$^{-1}$ on the red chip,  
with individual values reaching  $\sim$10--70 px$^{-1}$.
{ While, in similar programs of higher S/N, we were able to determine chemical abundance ratios of 
several $\alpha$- and heavy elements in GC stars \citep{Koch2017ESO}, this was not achievable from the present data set.}
\section{Radial velocities, cluster membership, and dynamical mass}
Radial velocities were measured via cross-correlation of the three strong CaT lines against a synthetic template of that region \citep{Kleyna2004}, 
which led to a median velocity error of 1.8 km\,s$^{-1}$. 
In Fig.~2 we show the run of velocities vs. radial distance from the nominal cluster center, here parameterized via it core radius. 
For clarity, this has been truncated at 10$\arcmin$ to emphasize the ``innermost'' region of our data set close to the actual cluster. 
\begin{figure}[b]
\centering
\includegraphics[width=1\hsize]{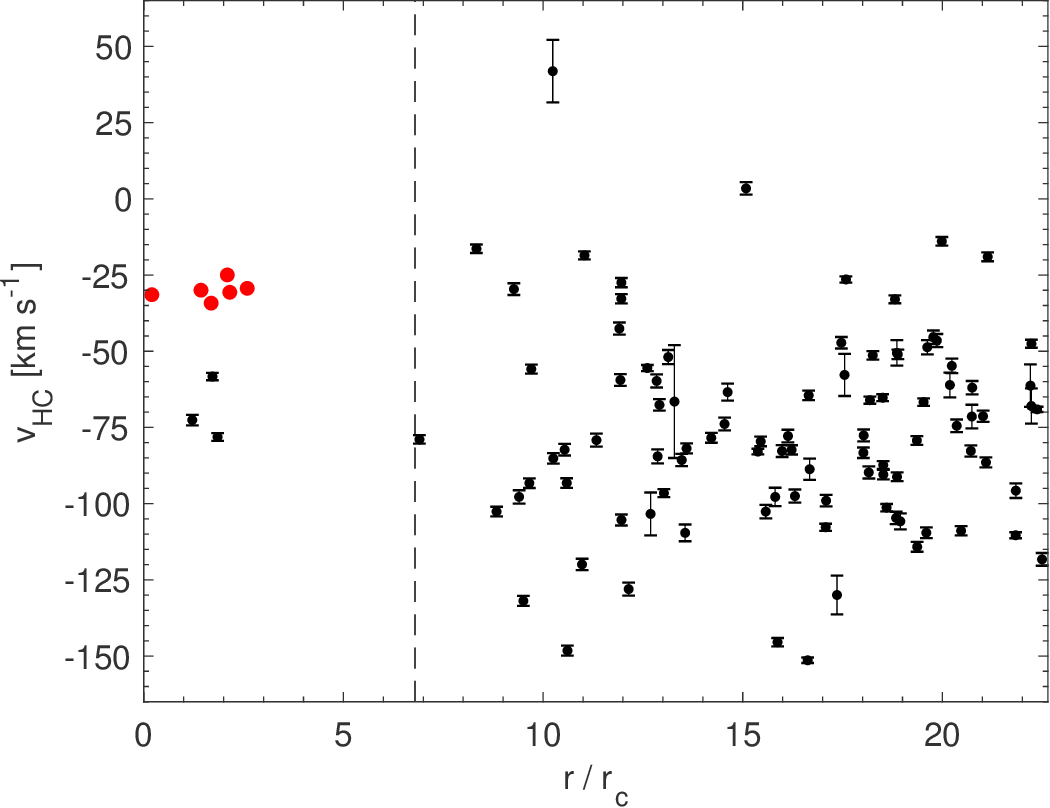}
\caption{Radial distribution of radial velocities in the inner sample, in units of the cluster's core-radius of 26.5$\arcsec$ \citep{Froebrich2008}. 
The vertical line illustrates the ``full extent'' of the cluster of 3$\arcmin$ as measured by \citet{Minniti2017}. Red
circles indicate the member candidates.}
\end{figure}

As this figure and the distribution of the entire sample (Fig.~3) show, isolating the signal from the cluster 
itself is severely hampered by the 
Galactic foreground contamination. 
{ The  mean  velocities  of these Galactic stars of $-61\pm2$ km\,s$^{-1}$ (heliocentric) and 
$-71\pm2$ km\,s$^{-1}$ (galactocentric) with a dispersion of 50$\pm$1 km\,s$^{-1}$ are 
fully compatible with the Galactic rotation curve towards this line of sight \citep[e.g.,][]{Dame2001}.}

We note, however, a group of six stars located within 2.6 core-radii of the center (red circles in Fig.~2)
that show very similar kinematic properties, clustering around a mean velocity of $-30$ km\,s$^{-1}$. Likewise, as we will show in Sect.~4, their the metallicities
are in excellent agreement with each other.
\begin{figure}[tb]
\centering
\includegraphics[width=1\hsize]{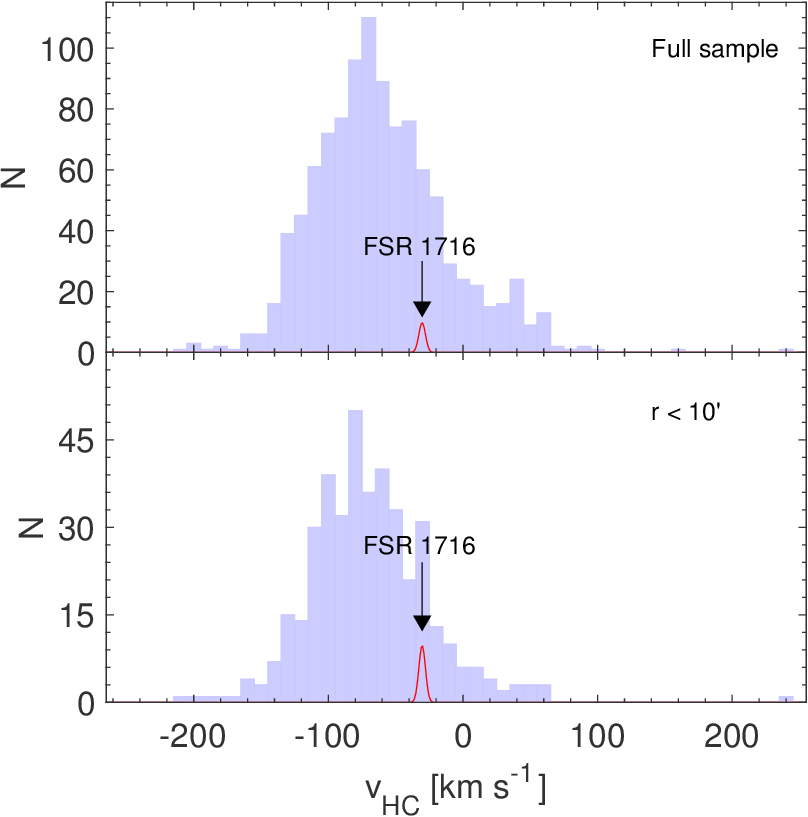}
\caption{Top panel: velocity histogram of the entire foreground sample including the six cluster member stars. 
The bottom panel only includes star with the inner 10$\arcmin$ to reflect the range shown in Fig.~2. 
The location of FSR~1716's mean velocity is each indicated by an arrow. 
}
\end{figure}
While the presence of a secondary plume of stars at $\sim -70$ km\,s$^{-1}$ could in principle be identified with an underlying cluster population as well, 
these stars have markedly different metallicities (Sect.~4) and we rather deem these foreground objects. 
The same holds for those stars outside of the cluster boundary of $\ga3\arcmin$, 
Thus we identify 
the innermost six stars at the higher velocity peak with members of FSR~1716.
Their individual properties are listed in Table~2.
\begin{table}[htb]
\caption{{ Derived properties} of the FSR~1716 member candidates}              
\centering          
\begin{tabular}{ccccc}     
\hline\hline       
  & S/N & v$_{\rm HC}$  & $\Sigma$ W & [Fe/H]$_{\rm CaT}$ \\
\raisebox{1.5ex}[-1.5ex]{Star} & [px$^{-1}$] & [km\,s$^{-1}$] & [\AA] & [dex] \\
\hline
{ Plate~1}  \phantom{1}\#16 & 26 & $-31.8\pm0.9$ & 4.40$\pm$0.11 & $-1.38\pm0.20$ \\
{ Plate~1}  \phantom{1}\#24 & 40 & $-31.0\pm1.6$ & 4.92$\pm$0.08 & $-1.28\pm0.20$ \\
{ Plate~1} \#106 & 20  & $-30.3\pm2.2$ & 4.19$\pm$0.15 & $-1.37\pm0.21$ \\
{ Plate~2} \#266 & 34 & $-25.3\pm1.2$ & 3.98$\pm$0.17 & $-1.51\pm0.21$ \\
{ Plate~3}  \phantom{1}\#76 & 28 & $-29.7\pm1.3$ & 4.98$\pm$0.11 & $-1.23\pm0.20$ \\
{ Plate~3} \#213 & 18 & $-34.6\pm1.8$ & 3.46$\pm$0.17 & $-1.64\pm0.21$ \\
\hline
\hline
\end{tabular}
\end{table}

Thereby, we establish the mean systemic velocity of FSR~1716 as $-$30.3$\pm$1.2 km\,s$^{-1}$ with a 
velocity dispersion of 2.5$\pm$0.9 km\,s$^{-1}$, a  
moderately low value typical of a low-to-intermediate mass Galactic star cluster \citep[e.g.,][]{Pryor1993,Koch2012,Koch2017Gaia}.
Using this information and the cluster morphology in terms of its characteristic \citet{King1966} radius 
\citep{Froebrich2008,BonattoBica2008,Minniti2017}, 
we can estimate the dynamical mass of FSR~1716 as $(1.4^{+1.2}_{-0.8})\times10^4$ M$_{\odot}$ \citep{Spitzer1987}, 
which is a factor of $\sim$70 larger than the low-mass estimate of \citet{BonattoBica2008}, albeit based on their premise of 
a significantly lower distance.  
Upon a typical stellar mass-to-light ratio of 1.5, this would place FSR~1716 at an approximate absolute magnitude of M$_V \sim -5.1\pm1.0$ mag, without doubt in the regimen of a GC. 

Ideally, the kinematic information should be coupled with accurate proper motions, allowing for an orbital computation that 
could hint at this cluster's origin. Unfortunately, none of the member stars had a cross match in the UCAC5 catalog 
\citep{Zacharias2017}, and only three stars could be identified in the HSOY (``Hot Stuff for One Year'') proper motion catalog \citep{Altmann2017}, which utilized the most recent positions from 
 Gaia-DR1 \citep{GaiaDR1}. However, the respective proper motions are highly inconclusive, leading to a broad variety 
 of possible orbits, from thick disk to outer halo; therefore, we do not pursue the orbital analyses any further. 

\section{Calcium triplet metallicities}
The three near-infrared CaT lines are a powerful indicator of stellar metallicities, [Fe/H]$_{\rm CaT}$, for RGB stars near and above the horizontal branch \citep{Armandroff1988}.
In the following, we adopted the calibrations of line strength with metallicity from \citet{Vasquez2015}, which are applicable into the metal-rich regime up to 0.7 dex, and which 
employ the infrared magnitude above the red clump ($K_S-K_{\rm RC}$). The latter is highly advantageous in population studies towards low latitudes, compared to 
the traditional use of the visual magnitude above the horizontal branch ($V - V_{\rm HB}$), which is more sensitive to reddening.  
In practice, these calibrations are parameterized as  
\begin{displaymath}
{\rm [Fe/H]_{\rm CaT}}\,=\,-3.15+0.432W' + 0.006W'^2, 
\end{displaymath}
with the reduced width $W'=\Sigma\,{\rm EW} + 0.384\,(K_S\,-\,K_{\rm RC})$. 
Here,  we defined the CaT line strength as the straight sum over the equivalent widths (EWs) of  the two strongest lines, i.e., $\Sigma\,{\rm EW} =  EW_{8542} + EW_{8662}$. 
The above formalism places our measurements on the metallicity scale of \citet{Carretta2009Fe}. 
The uncertainties on EW measurements and photometry were propagated through this process and added in quadrature to 
the R.M.S. scatter of 0.19 dex in the calibrations, as evaluated by \citet{Vasquez2015}. 
{ The latter component also overwhelmed the overall error budget compared to the photometric and EW uncertainties.}

We note that this procedure is explicitly only valid for stars above the red clump that are members to the system in question. 
Nominal metallicities assigned to foreground objects have no meaning, since their distance is {\em a priori}  unknown so that 
the formal magnitude above the clump, introduced as a proxy for absolute magnitude,  cannot be established. 
Furthermore, the CaT remains poorly measurable in many cold foreground stars in the sample 
 due to the lines lying within strong molecular bands \citep{Sharples1990,Kunder2012}. 

As Fig.~4, again truncated towards the inner 10$\arcmin$, shows, 
the velocity-selected member candidates share very similar CaT-metallicities, bolstering their identification with the cluster population.
\begin{figure}[!bt]
\centering
\includegraphics[width=1\hsize]{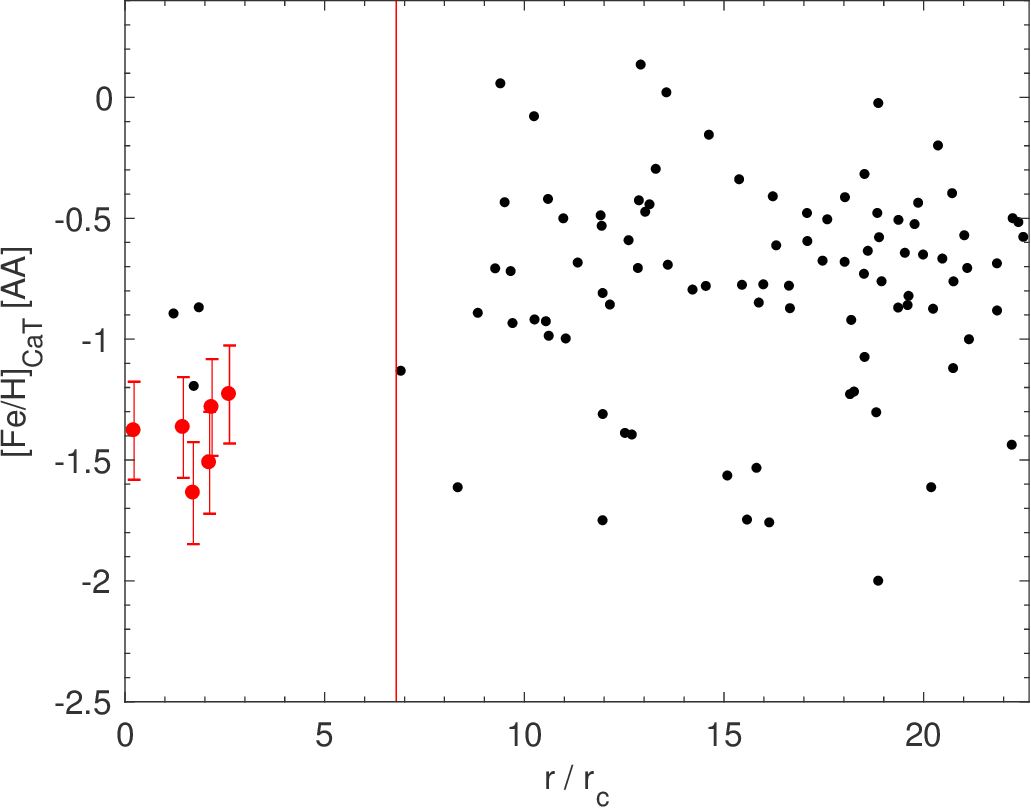}
\caption{CaT metallicities of the central parts of our observations. Member candidates are highlighted in red.}
\end{figure}
These stars are characterized by a mean
metallicity of $-1.38\pm0.20$ dex and a 1$\sigma$ dispersion of 0.15 dex  that is purely driven by the measurement errors that are typically of 0.20 dex.
This is in excellent agreement with previous measurements of the cluster mean from infrared photometry \citep[$-$1.6$^{+0.3}_{-0.6}$; ][]{Froebrich2008}. 
Likewise, \citet{Minniti2017} performed CMD-fitting on their VVV infrared photometry 
and used the period-metallicity relation for RR-Lyrae to ascertain a value of of $-$1.5$\pm$0.4 dex.
Combined with its kinematic properties and the appearance of its CMD, this argues in favour of this system indeed being an old, metal-poor GC, ruling out a younger open disk cluster or a Solar-metallicity object as suggested by \citet{Froebrich2008} 
and \citet{BonattoBica2008}. 
\section{Conclusions}
The low-latitude star cluster FSR~1716 has so far defied a clear-cut identification 
and has been placed in  various classes since its discovery. While originally suspected as an old open cluster at a close
distance and very low mass of only a few hundreds solar masses \citep{Froebrich2007FSR,BonattoBica2008}, 
the recent detection of old RR Lyrae stars associated with this object has argued in favour of a location near the 
Galactic bulge and an even older age, imprinting it with a globular label. 

The present spectroscopic study has firmly established FSR~1716 as a metal-poor system (at $-1.4$ dex), in line with photometric measurements. Our kinematic measurements confirmed a low mass on the order of 10$^4$ M$_{\odot}$, thus 
rendering this object an old, metal-poor and low-mass globular cluster that is currently located in the inner disk of the Milky Way.
Clearly, improved proper motions that can be gleaned from long-baseline, high-accuracy surveys 
\citep[e.g.,][]{Zacharias2017,Minniti2017} are needed to  aid in further investigation the history of this intriguing object.
\begin{acknowledgements}
We are grateful to the referee, Jeffrey Simpson, for a very swift and helpful report.
Based on data acquired through the Australian Astronomical Observatory, via program NOAO/0195  (PI: A. Kunder).
The research leading to these results has received funding from the European Community's Seventh Framework Programme (FP7/2013-2016) under grant agreement number 312430 (OPTICON, proposal 17A/051).
This work was supported in parts by Sonderforschungsbereich SFB 881 "The Milky Way System" (subproject A8) 
of the German Research Foundation (DFG).
\end{acknowledgements}
\bibliographystyle{aa} 
\bibliography{ms} 

\begin{thebibliography}{37}
\expandafter\ifx\csname natexlab\endcsname\relax\def\natexlab#1{#1}\fi

\bibitem[{{AAO software Team}(2015)}]{2dfdr}
{AAO software Team}. 2015, {2dfdr: Data reduction software}, Astrophysics
  Source Code Library

\bibitem[{{Altmann} {et~al.}(2017){Altmann}, {Roeser}, {Demleitner}, {Bastian},
  \& {Schilbach}}]{Altmann2017}
{Altmann}, M., {Roeser}, S., {Demleitner}, M., {Bastian}, U., \& {Schilbach},
  E. 2017, \aap, 600, L4

\bibitem[{{Armandroff} \& {Zinn}(1988)}]{Armandroff1988}
{Armandroff}, T.~E. \& {Zinn}, R. 1988, \aj, 96, 92

\bibitem[{{Bonatto} \& {Bica}(2007)}]{BonattoBica2007}
{Bonatto}, C. \& {Bica}, E. 2007, \aap, 473, 445

\bibitem[{{Bonatto} \& {Bica}(2008)}]{BonattoBica2008}
{Bonatto}, C. \& {Bica}, E. 2008, \aap, 491, 767

\bibitem[{{Brunthaler} {et~al.}(2011){Brunthaler}, {Reid}, {Menten}, {Zheng},
  {Bartkiewicz}, {Choi}, {Dame}, {Hachisuka}, {Immer}, {Moellenbrock},
  {Moscadelli}, {Rygl}, {Sanna}, {Sato}, {Wu}, {Xu}, \&
  {Zhang}}]{Brunthaler2011}
{Brunthaler}, A., {Reid}, M.~J., {Menten}, K.~M., {et~al.} 2011, Astronomische
  Nachrichten, 332, 461

\bibitem[{{Buckner} \& {Froebrich}(2016)}]{Buckner2016}
{Buckner}, A.~S.~M. \& {Froebrich}, D. 2016, ArXiv e-prints

\bibitem[{{Carretta} {et~al.}(2009){Carretta}, {Bragaglia}, {Gratton},
  {D'Orazi}, \& {Lucatello}}]{Carretta2009Fe}
{Carretta}, E., {Bragaglia}, A., {Gratton}, R., {D'Orazi}, V., \& {Lucatello},
  S. 2009, \aap, 508, 695

\bibitem[{{Cutri} {et~al.}(2003){Cutri}, {Skrutskie}, {van Dyk}, {Beichman},
  {Carpenter}, {Chester}, {Cambresy}, {Evans}, {Fowler}, {Gizis}, {Howard},
  {Huchra}, {Jarrett}, {Kopan}, {Kirkpatrick}, {Light}, {Marsh}, {McCallon},
  {Schneider}, {Stiening}, {Sykes}, {Weinberg}, {Wheaton}, {Wheelock}, \&
  {Zacarias}}]{Cutri2003}
{Cutri}, R.~M., {Skrutskie}, M.~F., {van Dyk}, S., {et~al.} 2003, {2MASS All
  Sky Catalog of point sources.}

\bibitem[{{Dame} {et~al.}(2001){Dame}, {Hartmann}, \& {Thaddeus}}]{Dame2001}
{Dame}, T.~M., {Hartmann}, D., \& {Thaddeus}, P. 2001, \apj, 547, 792

\bibitem[{{Dotter} {et~al.}(2008){Dotter}, {Chaboyer}, {Jevremovi{\'c}},
  {Kostov}, {Baron}, \& {Ferguson}}]{Dotter2008}
{Dotter}, A., {Chaboyer}, B., {Jevremovi{\'c}}, D., {et~al.} 2008, \apjs, 178,
  89

\bibitem[{{Friel}(2013)}]{Friel2013}
{Friel}, E.~D. 2013, {Open Clusters and Their Role in the Galaxy}, ed. T.~D.
  {Oswalt} \& G.~{Gilmore}, 347

\bibitem[{{Froebrich} {et~al.}(2008){Froebrich}, {Meusinger}, \&
  {Scholz}}]{Froebrich2008}
{Froebrich}, D., {Meusinger}, H., \& {Scholz}, A. 2008, \mnras, 390, 1598

\bibitem[{{Froebrich} {et~al.}(2007){Froebrich}, {Scholz}, \&
  {Raftery}}]{Froebrich2007FSR}
{Froebrich}, D., {Scholz}, A., \& {Raftery}, C.~L. 2007, \mnras, 374, 399

\bibitem[{{Ivanov} {et~al.}(2005){Ivanov}, {Kurtev}, \&
  {Borissova}}]{Ivanov2005}
{Ivanov}, V.~D., {Kurtev}, R., \& {Borissova}, J. 2005, \aap, 442, 195

\bibitem[{{King}(1966)}]{King1966}
{King}, I.~R. 1966, \aj, 71, 64

\bibitem[{{Kleyna} {et~al.}(2004){Kleyna}, {Wilkinson}, {Evans}, \&
  {Gilmore}}]{Kleyna2004}
{Kleyna}, J.~T., {Wilkinson}, M.~I., {Evans}, N.~W., \& {Gilmore}, G. 2004,
  \mnras, 354, L66

\bibitem[{{Koch} {et~al.}(2017{\natexlab{a}}){Koch}, {Hansen}, \&
  {Kunder}}]{Koch2017ESO}
{Koch}, A., {Hansen}, C.~J., \& {Kunder}, A. 2017{\natexlab{a}}, \aap, 604, A41

\bibitem[{{Koch} {et~al.}(2017{\natexlab{b}}){Koch}, {Hansen}, \&
  {Kunder}}]{Koch2017Gaia}
{Koch}, A., {Hansen}, T.~T., \& {Kunder}, A. 2017{\natexlab{b}}, \aap,
  submitted

\bibitem[{{Koch} {et~al.}(2012){Koch}, {L{\'e}pine}, \& {{\c C}al{\i}{\c
  s}kan}}]{Koch2012}
{Koch}, A., {L{\'e}pine}, S., \& {{\c C}al{\i}{\c s}kan}, {\c S}. 2012, in
  European Physical Journal Web of Conferences, Vol.~19, European Physical
  Journal Web of Conferences, 03002

\bibitem[{{Koch} {et~al.}(2016){Koch}, {McWilliam}, {Preston}, \&
  {Thompson}}]{Koch2016}
{Koch}, A., {McWilliam}, A., {Preston}, G.~W., \& {Thompson}, I.~B. 2016, \aap,
  587, A124

\bibitem[{{Kunder} {et~al.}(2012){Kunder}, {Koch}, {Rich}, {de Propris},
  {Howard}, {Stubbs}, {Johnson}, {Shen}, {Wang}, {Robin}, {Kormendy}, {Soto},
  {Frinchaboy}, {Reitzel}, {Zhao}, \& {Origlia}}]{Kunder2012}
{Kunder}, A., {Koch}, A., {Rich}, R.~M., {et~al.} 2012, \aj, 143, 57

\bibitem[{{Lamers} {et~al.}(2005){Lamers}, {Gieles}, {Bastian}, {Baumgardt},
  {Kharchenko}, \& {Portegies Zwart}}]{Lamers2005}
{Lamers}, H.~J.~G.~L.~M., {Gieles}, M., {Bastian}, N., {et~al.} 2005, \aap,
  441, 117

\bibitem[{{Law} \& {Majewski}(2010)}]{Law2010}
{Law}, D.~R. \& {Majewski}, S.~R. 2010, \apj, 718, 1128

\bibitem[{{Mar{\'{\i}}n-Franch} {et~al.}(2009){Mar{\'{\i}}n-Franch},
  {Aparicio}, {Piotto}, {Rosenberg}, {Chaboyer}, {Sarajedini}, {Siegel},
  {Anderson}, {Bedin}, {Dotter}, {Hempel}, {King}, {Majewski}, {Milone},
  {Paust}, \& {Reid}}]{Marin-Franch2009}
{Mar{\'{\i}}n-Franch}, A., {Aparicio}, A., {Piotto}, G., {et~al.} 2009, \apj,
  694, 1498

\bibitem[{{Minniti} {et~al.}(2017){Minniti}, {Palma}, {D{\'e}k{\'a}ny},
  {Hempel}, {Rejkuba}, {Pullen}, {Alonso-Garc{\'{\i}}a}, {Barb{\'a}}, {Barbuy},
  {Bica}, {Bonatto}, {Borissova}, {Catelan}, {Carballo-Bello}, {Chene},
  {Clari{\'a}}, {Cohen}, {Contreras Ramos}, {Dias}, {Emerson}, {Froebrich},
  {Buckner}, {Geisler}, {Gonzalez}, {Gran}, {Hajdu}, {Irwin}, {Ivanov},
  {Kurtev}, {Lucas}, {Majaess}, {Mauro}, {Moni-Bidin}, {Navarrete},
  {Ram{\'{\i}}rez Alegr{\'{\i}}a}, {Saito}, {Valenti}, \&
  {Zoccali}}]{Minniti2017}
{Minniti}, D., {Palma}, T., {D{\'e}k{\'a}ny}, I., {et~al.} 2017, \apjl, 838,
  L14

\bibitem[{{Mucciarelli} {et~al.}(2017){Mucciarelli}, {Monaco}, {Bonifacio}, \&
  {Saviane}}]{Mucciarelli2017}
{Mucciarelli}, A., {Monaco}, L., {Bonifacio}, P., \& {Saviane}, I. 2017, \aap,
  603, L7

\bibitem[{{Prusti} {et~al.}(2016){Prusti}, {de Bruijne}, {Brown}, {Vallenari},
  {Babusiaux}, {Bailer-Jones}, {Bastian}, {Biermann}, {Evans}, \&
  et~al.}]{GaiaDR1}
{Prusti}, T., {de Bruijne}, J.~H.~J., {Brown}, A.~G.~A., {et~al.} 2016, \aap,
  595, A1

\bibitem[{{Pryor} \& {Meylan}(1993)}]{Pryor1993}
{Pryor}, C. \& {Meylan}, G. 1993, in Astronomical Society of the Pacific
  Conference Series, Vol.~50, Structure and Dynamics of Globular Clusters, ed.
  S.~G. {Djorgovski} \& G.~{Meylan}, 357

\bibitem[{{Recio-Blanco} {et~al.}(2014){Recio-Blanco}, {de Laverny},
  {Kordopatis}, {Helmi}, {Hill}, {Gilmore}, {Wyse}, {Adibekyan}, {Randich},
  {Asplund}, {Feltzing}, {Jeffries}, {Micela}, {Vallenari}, {Alfaro}, {Allende
  Prieto}, {Bensby}, {Bragaglia}, {Flaccomio}, {Koposov}, {Korn}, {Lanzafame},
  {Pancino}, {Smiljanic}, {Jackson}, {Lewis}, {Magrini}, {Morbidelli},
  {Prisinzano}, {Sacco}, {Worley}, {Hourihane}, {Bergemann}, {Costado},
  {Heiter}, {Joffre}, {Lardo}, {Lind}, \& {Maiorca}}]{RecioBlanco2014}
{Recio-Blanco}, A., {de Laverny}, P., {Kordopatis}, G., {et~al.} 2014, \aap,
  567, A5

\bibitem[{{Reid} {et~al.}(2009){Reid}, {Menten}, {Zheng}, {Brunthaler},
  {Moscadelli}, {Xu}, {Zhang}, {Sato}, {Honma}, {Hirota}, {Hachisuka}, {Choi},
  {Moellenbrock}, \& {Bartkiewicz}}]{Reid2009}
{Reid}, M.~J., {Menten}, K.~M., {Zheng}, X.~W., {et~al.} 2009, \apj, 700, 137

\bibitem[{{Sharp} {et~al.}(2006){Sharp}, {Saunders}, {Smith}, {Churilov},
  {Correll}, {Dawson}, {Farrel}, {Frost}, {Haynes}, {Heald}, {Lankshear},
  {Mayfield}, {Waller}, \& {Whittard}}]{Sharp2006}
{Sharp}, R., {Saunders}, W., {Smith}, G., {et~al.} 2006, in \procspie, Vol.
  6269, Society of Photo-Optical Instrumentation Engineers (SPIE) Conference
  Series, 62690G

\bibitem[{{Sharples} {et~al.}(1990){Sharples}, {Walker}, \&
  {Cropper}}]{Sharples1990}
{Sharples}, R., {Walker}, A., \& {Cropper}, M. 1990, \mnras, 246, 54

\bibitem[{{Spitzer}(1987)}]{Spitzer1987}
{Spitzer}, L. 1987, {Dynamical evolution of globular clusters}

\bibitem[{{Vande Putte} {et~al.}(2010){Vande Putte}, {Garnier}, {Ferreras},
  {Mignani}, \& {Cropper}}]{VandePutte2010}
{Vande Putte}, D., {Garnier}, T.~P., {Ferreras}, I., {Mignani}, R.~P., \&
  {Cropper}, M. 2010, \mnras, 407, 2109

\bibitem[{{V{\'a}squez} {et~al.}(2015){V{\'a}squez}, {Zoccali}, {Hill},
  {Gonzalez}, {Saviane}, {Rejkuba}, \& {Battaglia}}]{Vasquez2015}
{V{\'a}squez}, S., {Zoccali}, M., {Hill}, V., {et~al.} 2015, \aap, 580, A121

\bibitem[{{Zacharias} {et~al.}(2017){Zacharias}, {Finch}, \&
  {Frouard}}]{Zacharias2017}
{Zacharias}, N., {Finch}, C., \& {Frouard}, J. 2017, \aj, 153, 166

\end{thebibliography}
\end{document}